\documentstyle[12pt]{article}
\begin{document}
\title{Wave functions and tunneling times for one-dimensional
transmission and reflection.}
\author{N L Chuprikov \\ Tomsk State Pedagogical University \\ 634041
Tomsk, Russia \\ E-mail: chnl@tspu.edu.ru}

\maketitle

\begin{abstract}

It is shown that in the case of the one-particle one-dimensional
scattering problem for a given time-independent potential, for each
state of the whole quantum ensemble of identically prepared particles,
there is an unique pair of (subensemble's) solutions to the
Schr\"odinger equation, which, as we postulate, describe separately
transmission and reflection: in the case of nonstationary states, for
any instant of time, these functions are orthogonal and their sum
describes the state of all particles; evolving with constant norms, one
of them approaches at late times the transmitted wave packet and
another approaches the reflected packet. Both for transmission and
reflection, 1) well before and after the scattering event, the average
kinetic energy of particles is the same, 2) the average starting point
differs, in the general case, from that for all particles. It is shown
that for reflection, in the case of symmetric potential barriers, the
domain of the motion of particles is bounded by the midpoint of the
barrier region. We define (exact and asymptotic) transmission and
reflection times and show that the basic results of our formalism can
be, in principle, checked experimentally.

\vspace{1cm} PACS numbers:  03.65.Ca,03.65.Xp

\end{abstract}
\newpage

\newcommand{\Api}{A_{in}}
\newcommand{\Ami}{B_{in}}
\newcommand{\Apo}{A_{out}}
\newcommand{\Amo}{B_{out}}
\newcommand{\bpi}{a_{in}}
\newcommand{\bmi}{b_{in}}
\newcommand{\bpo}{a_{out}}
\newcommand{\bmo}{b_{out}}
\newcommand{\api}{a_{in}}
\newcommand{\ami}{b_{in}}
\newcommand{\apo}{a_{out}}
\newcommand{\amo}{b_{out}}

\section{Introduction}

For a long time tunneling a particle through an one-dimensional
time-independent potential barrier was considered in quantum mechanics
as a representative of well-understood phenomena. However, now it has
been realized that this is not the case. The inherent to quantum theory
standard wave-packet analysis (SWPA) \cite{Ha2,Wig,Har,Ha1,Ter} (see
also \cite{Col}), in which the study of the temporal aspects of
tunneling is reduced to following the centers of "mass" (CMs) of wave
packets, does not provide a clear prescription both how to interpret
properly the scattering of finite in $x$ space wave packets and how to
introduce characteristic times for a {\it tunneling} particle. All
these questions constitute the main content of the so-called tunneling
time problem (TTP) which have been of great interest for the last
decades.

As is known, the main peculiarity of the tunneling of finite wave
packets is that the average particle's kinetic energy for the
transmitted, reflected and incident wave packets is different. For
example, in the case of tunneling through an opaque rectangular
barrier, the average velocity of the transmitted particle is larger
than that of the incident particle. It is evident that this fact needs
a proper explanation. As was pointed out in \cite{La2,La1}, it would be
strange to interpret the above property of wave packets as the evidence
of accelerating a particle (in the asymptotic regions) by the static
potential barrier.  Besides, in this case there is no causal link
between the transmitted and incident wave packets (see \cite{La2,La1}).

As regards wide (strictly speaking, infinite) in $x$ space wave
packets, the average kinetic energy of particles, before and after the
interaction, is the same. But now the uncertainty in defining the CM's
position and, consequently, corresponding asymptotic times is very
large. As a result, the most of physicists considers the characteristic
times introduced in the SWPA as quantities having no physical sense.
The review \cite{Ha2} devoted to the TTP seems to be the last one in
which the SWPA is considered in a positive context.

Apart from the SWPA, in the same or different setting the tunneling
problem, a variety of alternative approaches (see reviews
\cite{Ha2,La1,Olk,Ste,Mu0,Nu0} and references therein) to advance
different characteristic times for a tunneling particle have also been
developed. Among the alternative conceptions, of interest are that of
the dwell time \cite{Smi,Ja1,Le1,Nus}, that of the Larmor time
\cite{Baz,Ryb,But,Bu1,Zhi} to give the way of measuring the dwell time,
and the conception of the time of arrival which is based on introducing
either a suitable time operator (see, for example,
\cite{Aha,Mu4,Hah,Noh,Mu9}) or the positive operator valued measure
\cite{Mu0}. Besides, of interest are attempts to study the temporal
aspects of tunneling in the frame of the Feynman, Bohmian and Wigner
approaches to deal with the random trajectories of particles (see, for
example, \cite{Sok,Bo1,Yam,Ymm,Gru} and references therein). One should
also mention the papers \cite{Ga1,Ga2,Ga3} where the TTP is studied in
the framework of a nonstandard setting the scattering problem.
But again, for a particle whose initial state is described by a
Gaussian-like wave packet, none of the alternative approaches have not
yet led to commonly accepted characteristic times (see the reviews
\cite{Ha2,La1,Olk,Ste,Mu0,Nu0}).

Note, unlike the SWPA, in all these approaches theoretical efforts have
been aimed at elaborating some rule of timing the motion of a quantum
particle. As was said in \cite{Mu0}, "... up to now the interest of
theoreticians [to the TTP] has been motivated ...  by a fundamental
lacuna of quantum theory, namely the absence of a clear prescription to
incorporate time observables into its formalism".  However, in our
opinion, there is no necessity in such a prescription: time in quantum
mechanics has the status of a parameter, and, hence, there is no room
here for time observables (or, time operators). And, what is more
important, the rule of timing the particle's motion have already been
available in quantum theory, and this rule is dictated by the
correspondence principle.

By the analogue with classical mechanics where timing the particle's
motion is reduced to the analysis of the function $x(t)$ ($x$ is the
particle's position, $t$ is time), in quantum mechanics characteristic
times for a particle should be derived from studying the temporal
dependence of the expectation (average) value of the position operator
for a particle in a given state (or, what is equivalent, from studying
the temporal behavior of the CM of the corresponding wave packet).
Besides, from the analysis of the temporal dependence of the mean-square
deviation for this operator (or, from the analysis of the temporal
behavior of the corresponding leading and trailing edges of the wave
packet) one can take into account uncertainty in the particle's
position, and thereby evaluate the error of the above timing.

However, one has to bear in mind the following. The above timing
procedure suggests that the average value of the position operator has
its primary physical sense, as the most probable position of a
particle, at all stages of its motion. For a free particle whose state
is described by a Gaussian-like wave packet, this requirement is
fulfilled and, thus, no problem arises in timing its motion. An
essentially different situation takes place in the case of a {\it
tunneling} particle. Now, following the CM of the wave packet to
describe the state of the whole ensemble of particles becomes
meaningless at some stages of scattering. In particular, after the
scattering event, when we deal, in fact, with two scattered
(transmitted and reflected) wave packets, the averaging over the whole
ensemble of particles has no physical sense.  Of course, in this case
there is a possibility to define the individual average positions of
transmitted and reflected particles. However, in timing, such averaging
suggests the separate description of both the subensembles at early
times, what is widely accepted to be impossible in quantum mechanics.
As was pointed out in \cite{Nu0}, "...  transmission and reflection are
inextricably intertwined".

So, quantum theory provides the needed rule of timing the motion of a
particle, but it conflicts with the existing viewpoint that
transmission and reflection are allegedly inseparable (namely this
obstacle have remained to overcome in the SWPA). However, in our
opinion, there is no basis for such a viewpoint. For none of the
principles of quantum mechanics forbids such a separation. In reality,
the main problem is that quantum theory, as it stands, does not provide
the way of separating transmission and reflection. In our opinion, just
the absence of the corresponding mathematical formalism is a
fundamental lacuna in quantum theory.

In this paper, in the framework of the conventional quantum mechanics,
we show that, at least in one dimension, transmission and reflection
can be described separately. It is surprising that such a description
needs no innovation. As it has turned out, the separation of
transmission and reflection is provided by the intrinsic property of
the Schr\"odinger equation, which have been overlooked before. Namely,
we show that in the standard setting the one-dimensional scattering
problem for a given potential the Schr\"odinger equation possesses, in
addition to the solution to describe the state of the whole ensemble of
particle, an unique pair of other solutions which, as we postulate,
describe separately transmission and reflection. The basis for such a
postulate is that, for any instant of time, the subensemble's
nonstationary-state wave functions are mutually orthogonal and their sum
describes the state of the whole ensemble of particles; one of them
causally evolves into the transmitted wave packet, and another
approaches at late times the reflected one. The main peculiarity of
stationary-state wave functions for transmission and reflection is
that, for a given energy of particle, there is a point in the barrier
region where these everywhere continuous functions have discontinuous
first spatial derivatives. Nevertheless, this point is not a sink or
source of particles for each subensemble. So that the norms of the
corresponding wave packets are constant in time.

Note, at present there is a paradoxical situation. Although the
tunneling phenomenon have been known for a long time, the properties
of tunneling proper have remained, in fact, unstudied. The point is
that the "full" wave function to describe the state of a particle in
the one-dimensional scattering problem relates to all particles of the
quantum ensemble, rather than to transmitted particles only. In this
connection, we hope that the formalism presented here will be useful
for a deeper understanding of the tunneling process and, in
particular, Hartman effect \cite{Har} widely discussed in the
literature (see, for example, \cite{Mu6}).

The paper is organized as follows. In Section \ref{a1} we pose a
complete one-dimensional scattering problem for a particle, and display
explicitly shortcomings to arise in the SWPA in solving the TTP. In
Section \ref{a2} we present a renewed wave-packet analysis in which
transmission and reflection are treated separately. In Section \ref{a3}
we define the average (exact and asymptotic) transmission and
reflection times and consider, in details, the cases of rectangular
barriers and $\delta$-potentials.

\newcommand{\ko}{\kappa_0^2}
\newcommand{\kj}{\kappa_j^2}
\newcommand{\kd}{\kappa_j d_j}
\newcommand{\kki}{\kappa_0\kappa_j}

\newcommand{\Ra}{R_{j+1}}
\newcommand{\Rb}{R_{(1,j)}}
\newcommand{\Rc}{R_{(1,j+1)}}

\newcommand{\Ta}{T_{j+1}}
\newcommand{\Tb}{T_{(1,j)}}
\newcommand{\Tc}{T_{(1,j+1)}}

\newcommand{\Wa}{w_{j+1}}
\newcommand{\Wb}{w_{(1,j)}}
\newcommand{\Wc}{w_{(1,j+1)}}

\newcommand{\UU}{u^{(+)}_{(1,j)}}
\newcommand{\VV}{u^{(-)}_{(1,j)}}

\newcommand{\ta}{t_{j+1}}
\newcommand{\tb}{t_{(1,j)}}
\newcommand{\tc}{t_{(1,j+1)}}

\newcommand{\tee}{\vartheta_{(1,j)}}

\newcommand{\tta}{\tau_{j+1}}
\newcommand{\ttb}{\tau_{(1,j)}}
\newcommand{\ttc}{\tau_{(1,j+1)}}

\newcommand{\FF}{\chi_{(1,j)}}
\newcommand {\aro}{(k)}
\newcommand {\da}{\partial}
\newcommand{\ppp}{\mbox{\hspace{5mm}}}
\newcommand{\ooo}{\mbox{\hspace{3mm}}}
\newcommand{\ooa}{\mbox{\hspace{1mm}}}

\section{Setting the problem for a completed scattering} \label{a1}
\subsection{Backgrounds} \label{a11}

Let us consider a particle tunneling through the time-independent
potential barrier $V(x)$ confined to the finite spatial interval
$[a,b]$ $(a>0)$; $d=b-a$ is the barrier width. Let its in state,
$\Psi_{in}(x),$ at $t=0$ be the normalized function
$\Psi^{(0)}_{left}(x)$. The function is proposed to belong to the set
$S_{\infty}$ consisting from infinitely differentiable functions
vanishing exponentially in the limit $|x|\to \infty$. The
Fourier-transform of such functions are known to belong to the set
$S_{\infty}$ as well. In this case the position and momentum operators
both are well-defined. Without loss of generality we will suppose that

\[<\Psi^{(0)}_{left}|\hat{x}|\Psi^{(0)}_{left}>=0,\ppp
<\Psi^{(0)}_{left}|\hat{p}|\Psi^{(0)}_{left}> =\hbar k_0 > 0,\ppp
<\Psi^{(0)}_{left}|\hat{x}^2|\Psi^{(0)}_{left}>
=l_0^2,\]
here $l_0$ is the wave-packet's half-width at $t=0$ ($l_0<<a$);
$\hat{x}$ and $\hat{p}$ are the operators of the particle's position
and momentum, respectively.

An important restriction should be imposed also on the rate of spreading
the incident wave packet. Namely, we will suppose that the average
velocity $\hbar k_0/m$ is large enough, so that in the incident wave
packet its parts lying behind the CM within the wave-packet's half-width
move toward the barrier.

As is known, the formal solution to the temporal one-dimensional
Schr\"odinger equation (OSE) of the problem can be written as
$e^{-i\hat{H}t/\hbar}\Psi_{in}(x).$ To solve explicitly this equation,
we will use here the variant (see \cite{Ch1}) of the well-known
transfer matrix method \cite{Mez} that allows one to calculate the
tunneling parameters, as well as to connect the amplitudes of the
outgoing and corresponding incoming waves, for any system of potential
barriers. The state of a particle with the wave-number $k$ can be
written in the form

\begin{eqnarray} \label{1}
\Psi_{full}(x;k)=\Api(k)e^{ikx}+\Amo(k)e^{-ikx},
\end{eqnarray}
for $x\le a$, and

\begin{eqnarray} \label{2}
\Psi_{full}(x;k)=\Apo(k)e^{ikx} +\Ami(k)e^{-ikx},
\end{eqnarray}

\noindent for $x>b$, where $\Api(k)$ should be found from the initial
condition; $\Ami(k)=0.$ The coefficients entering this solution are
connected by the transfer matrix ${\bf Y}$:

\begin{eqnarray} \label{50}
\left(\begin{array}{c} \Api \\ \Amo
\end{array} \right)={\bf Y} \left(\begin{array}{c} \Apo \\ \Ami
\end{array} \right); \hspace{8mm}
{\bf Y}=\left(\begin{array}{cc} q & p \\ p^* &
q^* \end{array} \right);
\end{eqnarray}
which can be expressed in terms of the real tunneling parameters $T$,
$J$ and $F$,

\newcommand{\iii}{\mbox{\hspace{10mm}}}

\begin{eqnarray} \label{500}
q=\frac{1}{\sqrt{T(k)}}\exp\left[i(kd-J(k))\right];\ppp
p=\sqrt{\frac{R(k)}{T(k)}}\exp\left[i\left(\frac{\pi}{2}+
F(k)-ks\right)\right];
\end{eqnarray}

\noindent $T(k)$  (the real transmission coefficient) and $J(k)$
(phase) are even and odd functions of $k$, respectively;
$F(-k)=\pi-F(k)$; $R(k)=1-T(k)$; $s=a+b$. Note that the functions
$T(k)$, $J(k)$ and $F(k)$ contain all needed information about the
influence of the potential barrier on a particle. We will suppose that
the tunneling parameters have already been known explicitly. To find
them, one can use the recurrence relations obtained in \cite{Ch1} just
for these real parameters.

As is known, solving the TTP is reduced in the SWPA to timing a
particle beyond the scattering region where the exact solution of the
OSE approaches the corresponding in or out asymptote \cite{Tei}.
Thus, definitions of characteristic times in this approach can be done
in terms of the in and out asymptotes of the tunneling problem.

Note that in asymptote in the one-dimensional scattering problem
represents an one-packet object to converge, well before the
scattering event, with the incident wave packet, while out asymptote
represents the superposition of two non-overlapped wave packets to
converge, at $t\to \infty,$ with the transmitted and reflected ones.
It is easy to show that in the problem at hand in asymptote,
$\Psi_{in}(x,t)$, and out asymptote, $\Psi_{out}(x,t)$, can be written
as follows

\begin{eqnarray} \label{59}
\Psi_{in}(x,t)=\frac{1}{\sqrt{2\pi}}\int_{-\infty}^{\infty}
f_{in}(k,t) e^{ikx}dk,\ppp f_{in}(k,t)=\Api(k) \exp[-i E(k)t/\hbar];
\end{eqnarray}

\begin{eqnarray} \label{60}
\Psi_{out}(x,t)=\frac{1}{\sqrt{2\pi}}\int_{-\infty}^{\infty}
\left[f_{tr}(k,t)+f_{ref}(k,t)\right] e^{ikx}dk,\ooo f_{out}(k,t)=
f_{out}^{tr}(k,t)+f_{out}^{ref}(k,t);
\end{eqnarray}

\begin{eqnarray} \label{61}
f_{out}^{tr}(k,t)=\sqrt{T(k)}\Api(k)
\exp[i(J(k)-kd -E(k)t/\hbar)];
\end{eqnarray}

\begin{eqnarray} \label{62}
f_{out}^{ref}(k,t)=\sqrt{R(k)}\Api(-k)
\exp[-i(J(k)-F(k)-\frac{\pi}{2}+2ka+E(k)t/\hbar)]
\end{eqnarray}
where $E(k)=\hbar^2 k^2/2m.$

For a completed scattering we have
\[\Psi_{full}(x,t)\approx\Psi_{in}(x,t)\ppp \mbox{when}\ppp t=0,\]
\[\Psi_{full}(x,t)=\Psi_{out}(x,t) \ppp\mbox{when}\ppp t\to\infty.\]
It is obvious that the larger is the distance $a$, the more correct is
the approximation for $\Psi_{full}(x,t)$ at $t=0$.

For particles starting (on the average) from the origin, we have
\begin{eqnarray} \label{63}
<\hat{x}>_{in}=\frac{\hbar k_0}{m}t
\end{eqnarray} (hereinafter, for any Hermitian operator $\hat{Q}$
\[<\hat{Q}>_{in}=\frac{<f_{in}|\hat{Q}|f_{in}>}{<f_{in}|f_{in}>};\]
similar notations are used below for the transmitted and reflected
wave packets). The averaging separately over the transmitted and
reflected wave packets yields

\begin{eqnarray} \label{64}
<\hat{x}>^{tr}_{out}=\frac{\hbar t}{m}<k>^{tr}_{out}
-<J^\prime(k)>^{tr}_{out}+d;
\end{eqnarray}
\begin{eqnarray} \label{65}
<\hat{x}>^{ref}_{out}=\frac{\hbar t}{m}<k>^{ref}_{out}
+<J^\prime(k)-F^\prime(k)>^{ref}_{out}+2a
\end{eqnarray}
(hereinafter the prime denotes the derivative with respect to $k$).
Exps. (\ref{63}) --- (\ref{65}) yield the basis for defining the
asymptotic tunneling times in the SWPA.

\subsection {Problems of the standard wave-packet analysis} \label{a12}

To display explicitly some shortcomings of the SWPA, let us derive
again the SWPA's tunneling times. Let $Z_1$ be a point to lie at some
distance $L_1$ ($L_1\gg l_0$ and $a-L_1\gg l_0$) from the left boundary
of the barrier, and $Z_2$ be a point to lie at some distance $L_2$
($L_2\gg l_0$) from its right boundary. Following \cite{Ha1}, let us
define the difference between the times of arrival of the CMs of the
incident and transmitted packets at the points $Z_1$ and $Z_2$,
respectively (this time will be called below as the "transmission
time").  Analogously, let the "reflection time" be the difference
between the times of arrival of the CMs of the incident and reflected
packets at the same point $Z_1$.

Thus, let $t_1$ and $t_2$ be such instants of time that

\begin{equation} \label{8}
<\hat{x}>_{in}(t_1)=a-L_1; \ppp
<\hat{x}>^{tr}_{out}(t_2)=b+L_2.
\end{equation}

\noindent Then, considering (\ref{63}) and (\ref{64}), one can write the
"transmission time" $\Delta t_{tr}$ ($\Delta t_{tr} =t_2 -t_1$) for the
given interval in the form

\begin{eqnarray} \label{9}
\Delta t_{tr}=\frac{m}{\hbar}\Bigg[\frac{<J^\prime>_{out}^{tr} +L_2}
{<k>_{out}^{tr}} +\frac{L_1}{k_0}
+a\left(\frac{1}{<k>_{out}^{tr}} -\frac{1}{k_0}\right)\Bigg].
\end{eqnarray}
Similarly, for the reflected packet, let $t^{\prime}_1$ and
$t^{\prime}_2$ be such instants of time that

\begin{equation} \label{10}
<\hat{x}>_{in}(t^{\prime}_1)
=<\hat{x}>_{out}^{ref}(t^{\prime}_2)=a-L_1.
\end{equation}

\noindent From equations (\ref{63}), (\ref{65}) and (\ref{10}) it
follows that the "reflection time" $\Delta t_{ref}$ ($\Delta
t_{ref}=t^{\prime}_2-t^{\prime}_1$) can be written as

\begin{eqnarray} \label{11}
\Delta t_{ref}=\frac{m}{\hbar}\Bigg[\frac{<J^\prime -
F^\prime>_{out}^{ref} +L_1} {<-k>_{out}^{ref}} +\frac{L_1}{k_0}
+a\left(\frac{1}{<-k>_{out}^{ref}}
-\frac{1}{k_0}\right)\Bigg].
\end{eqnarray}

Note that the expectation values of $k$ for all three wave packets
coincide only in the limit $l_0\to\infty$ (i.e., for particles with a
well-defined momentum). In the general case these quantities are
distinguished. For example, for a particle whose initial state is
described by the Gaussian wave packet, we have

\[\Api(k)=A \exp(-l_0^2(k-k_0)^2), \ppp
A=\left(\frac{l_0^2}{\pi}\right)^{1/4}.\]

On can show that in this case
\begin{equation} \label{100}
<k>_{tr}=k_0+\frac{<T^\prime>_{in}} {4l_0^2<T>_{in}};
\end{equation}

\begin{equation} \label{101}
<-k>_{ref}=k_0+\frac{<R^\prime>_{in}} {4l_0^2<R>_{in}}.
\end{equation}
Let
\[<k>_{tr}=k_0+(\Delta k)_{tr},\ppp <-k>_{ref}=k_0+(\Delta k)_{ref},\]

\noindent then relations (\ref{100}) and (\ref{101}) can be written in
the form

\begin{equation} \label{102}
<T>_{in}\cdot (\Delta k)_{tr}=-<R>_{in}\cdot(\Delta k)_{ref}
=\frac{<T^\prime>_{in}}{4l_0^2}.
\end{equation}
\noindent Note that $R^\prime=-T^\prime$.

As is seen, quantities (\ref{9}) and (\ref{11}) cannot serve as
characteristic times for a particle.  Due to the last terms in these
expressions the above times depend essentially on the initial distance
between the wave packet and barrier, with $L_1$ being fixed.  These
terms are dominant for the sufficiently large distance $a$.  Moreover,
one of them must be negative. For example, for the transmitted wave
packet it takes place in the case of the under-barrier tunneling
through an opaque rectangular barrier. The numerical modelling of
tunneling \cite{Ha2,Ha1,Ter,Le1} shows in this case a premature
appearance of the CM of the transmitted packet behind the barrier, what
points to the lack of a causal link between the transmitted and
incident wave packets (see \cite{La1}).

As was shown in \cite{Ha2,Ha1}, this effect disappears in the limiting
case $l_0\to\infty$. For example, in the case of Gaussian wave packets
the fact that the last terms in (\ref{9}) and (\ref{11}) tend to zero
when $l_0\to \infty$, with the ratio $l_0/a$ being fixed, can be proved
with help of Exps. (\ref{100}) and (\ref{101}) (note that the limit
$l_0\to \infty$ with a fixed value of $a$ is unacceptable in this
analysis, because it contradicts the initial condition $a\gg l_0$ for a
completed scattering). Thus, at first glance, in the limit
$l_0\to\infty$ the SWPA seems to provide correct characteristic times
for a particle.  However, as will be seen from our formalism, even in
this case, the above times are poorly defined. The point is that the
above definitions of tunneling times for transmission and reflection
are based on the implicit assumption that particles of both the
subensembles start, on the average, from the origin as those of a whole
quantum ensemble. As will be seen from the following, this is not the
case.

Note, the fact that Exps. (\ref{9}) and (\ref{11}) cannot be applied to
particles does not at all mean that they are erroneous. These
expressions correctly describe the relative motion of the transmitted
(or reflected) and incident wave packets. The principal shortcoming of
the above approach is that it is meaningless to compare the motion of
the transmitted (or reflected) wave packet with that of the incident
one since they are related to different ensembles of particles and,
as a consequence, there is no causal relationship between them. The
right procedure of a separate timing of transmitted particles suggests
the availability of such initial wave packet to evolve causally into
the transmitted one.

\section {Separate description of transmission and reflection in the
one-dimensional scattering problem}\label{a2}

\hspace*{\parindent} For a long time the processes of transmission and
reflection in a quantum scattering have been accepted to be
inseparable, in principle. However, in this paper we show that, at
least in the one-dimensional case, there are sound reasons to consider
these processes separately.

\subsection {Wave function of a tunneling particle as a sum of wave
functions to describe separately transmission and reflection}
\label{a21}

According to quantum scattering theory, in one dimension,
stationary-state wave functions for a particle impinging the barrier
from the left (or from the right) possess one incoming and two outgoing
waves.  That is, in this case we deal, in fact, with one-source
two-sinks scattering problems.

Let for the problem at hand the amplitude of incoming wave, $\api$, be
equal to unit, then the amplitudes of all four waves read as

\begin{eqnarray} \label{700}
\api=1, \ppp \amo=\frac{p^*}{q}, \ppp \apo=\frac{1}{q}, \ppp \ami=0
\end{eqnarray}
(note, $\Amo=\amo \Api$, $\Apo=\apo \Api$, $\Ami=\ami \Api$).  Let us
also consider two auxiliary (two-sources one-sink) scattering tasks in
which the amplitudes of incoming and outgoing waves are

\begin{eqnarray} \label{701}
\bpi^{ref}=\frac{|p|^2}{|q|^2},\ppp\bmo^{ref}=\frac{p^*}{q},\ppp
\bpo^{ref}=0,\ppp\bmi^{ref}=\frac{p^*}{|q|^2};
\end{eqnarray}
and

\begin{eqnarray} \label{702}
\bpi^{tr}=\frac{1}{|q|^2},\ppp\bmo^{tr}=0,\ppp
\bpo^{tr}=\frac{1}{q},\ppp\bmi^{tr}=-\frac{p^*}{|q|^2}
\end{eqnarray}
(the transfer matrix (\ref{50}) is common for all three tasks).

Note that in the first auxiliary task the only outgoing wave coincides
with the reflected wave arising in the problem at hand (see
(\ref{700})). And, in the second task, the only outgoing wave coincides
with the transmitted wave in (\ref{700}). It is evident that the
sum of these two functions results just in that to describe the state
of a particle in the initial tunneling problem.

As is seen, the main peculiarity of the superposition of these two
states is that due to interference the incoming waves, in the region
$x>b$, disappear entirely (note that in the corresponding reverse
motion they are outgoing waves). Figuratively speaking, interference
reorients these waves into the region $x<a$. That is, in this
superposition the probability fields of both sinks are radically
reconstructed due to interference. Namely, they transform into fields
with one outgoing and one incoming waves.

Hereinafter, the wave function in which an incoming wave is associated
with the reflected wave of solution (\ref{700}) will be refereed to
as the reflection wave function (RWF).  Similarly, the wave function in
which an incoming wave is related to the transmitted wave of
(\ref{700}) will be refereed to as the transmission wave function
(TWF). We postulate that, in the considered scattering problem, namely
the nonstationary-state TWF and RWF describe, respectively, the
transmission and reflection processes. As will be shown below, both the
functions evolve in time with constant norms; at late times the TWF
(RWF) coincides with the transmitted (reflected) wave packet.

Thus, we see that the sum of wave functions (\ref{701}) and (\ref{702})
can be presented as that of the stationary-state RWF and TWF. Under the
reconstruction of the probability fields, the squared amplitude of the
incoming wave (in the region $x<a$) associated with reflection
increases due to interference from the initial value $|\bpi^{ref}|^2$
($=R^2$) (see (\ref{701})) to $|\bpi^{ref}|^2+|\bmi^{ref}|^2$
($=R^2+TR=R)$ (in the RWF). In the case of transmission the
corresponding quantity increases from the initial value $|\bpi^{tr}|^2$
($=T^2$) (see (\ref{702})) to $|\bpi^{tr}|^2+|\bmi^{tr}|^2$ ($=T^2+T
R=T)$ (in the TWF).

Of course, the above postulate suggests the availability of a proper
pair of solutions to the Schr\"odinger equation. The main thing which
should be taken into account in finding these solutions is that the RWF
describes the states of reflected particles only, and the TWF relates
only to transmitted particles. As was said above, in both the cases,
stationary solutions should contain one incoming and one outgoing wave.
In this paper we show that such solutions do exist.

\subsection {Wave functions for one-dimensional transmission and
reflection}\label{a22}

So, let $\Psi_{tr}$ and $\Psi_{ref}$ be the searched-for TWF and RWF,
respectively. In line with subsection \ref{a21}, their sum represents
the wave function to describe, in the problem at hand, the sate of the
whole ensemble of particles. Hence, from the mathematical point of view
our task now is to find such solutions $\Psi_{tr}$ and $\Psi_{ref}$ to
the Schr\"odinger equation that for any $t$,

\begin{equation} \label{261}
\Psi_{full}(x,t)=\Psi_{tr}(x,t)+\Psi_{ref}(x,t)
\end{equation}
where $\Psi_{full}(x,t)$ is the full wave function to describe all
particles (see section \ref{a1}). In the limit $t\to \infty$
\begin{equation} \label{262}
\Psi_{tr}(x,t)=\Psi_{out}^{tr}(x,t); \ppp
\Psi_{ref}(x,t)=\Psi_{out}^{ref}(x,t)
\end{equation}
where $\Psi_{out}^{tr}(x,t)$ and $\Psi_{out}^{ref}(x,t)$ are the
transmitted and reflected wave packets whose Fourier-transforms
presented in (\ref{61}) and (\ref{62}).

As is known, searching for the wave functions in the case of the
time-independent potential $V(x)$ is reduced to the solution of the
corresponding stationary Schr\"odinger equation. For a given $k$, let
us find firstly the functions $\Psi_{ref}(x;k)$ and $\Psi_{tr}(x;k)$
for the spatial region $x\le a$. In this region let

\begin{eqnarray} \label{265}
\Psi_{ref}(x;k)=\Api(k)\left(\Api^{ref}(k)e^{ikx}
+\Amo^{ref}(k)e^{-ikx}\right)
\end{eqnarray}
\begin{eqnarray} \label{2650}
\Psi_{tr}(x;k)=\Api(k)\Big(\Api^{tr}(k)e^{ikx}+
\Amo^{tr}(k)e^{-ikx}\Big)
\end{eqnarray}
where $\Api^{tr}+\Api^{ref}=1$, $\Amo^{tr}+\Amo^{ref}=\amo$.

Since the RWF describes the state of reflected particles only, the
probability flux for $\Psi_{ref}(x;k)$ should be equal to zero, i.e.,

\begin{eqnarray} \label{264}
|\Api^{ref}|^2-|\Amo^{ref}|^2=0.
\end{eqnarray}
In its turn, for $\Psi_{tr}(x;k)$ we have
\begin{eqnarray} \label{263}
|\Api^{tr}|^2-|\Amo^{tr}|^2=\frac{\hbar k}{m}T(k)
\end{eqnarray}
(the probability flux for the full wave function $\Psi_{full}(x;k)$ and
for $\Psi_{tr}(x;k)$ should be the same).

Taking into account that $\Psi_{tr}=\Psi_{full}-\Psi_{ref}$ let us now
exclude $\Psi_{tr}$ from Eq. (\ref{263}). As a result, we obtain for
$\Psi_{ref}$ the equation

\begin{eqnarray} \label{2630}
Re\left(\Api^{ref}\api^*-\Amo^{ref}\amo^* \right)=0.
\end{eqnarray}
The physical meaning of Eq. (\ref{2630}) is that the function
$\Psi_{ref}(x)$, with zero probability flux, is such that the sum of
the stationary-state RWF and any other stationary-state wave function
with a nonzero probability flux does not change the value of the
latter.

From condition (\ref{262}) for $\Psi_{ref}(x;k)$ it follows that
$\Amo^{ref}(k)=\amo(k)\equiv p^*/q$ (see (\ref{700})). Then Eq.
(\ref{2630}) yields $Re(\Api^{ref})=R$, and Eq. (\ref{264}) leads to
$|\Api^{ref}|^2= |\Amo^{ref}|^2=|p^*/q|^2=R.$ Thus,
$\Api^{ref}=\sqrt{R}(\sqrt{R}\pm i\sqrt{T}) \equiv
\sqrt{R}\exp(i\lambda)$; $\lambda=\pm\arctan(\sqrt{T/R})$.

So, there are two solutions to satisfy the above requirements for
$\Psi_{ref}(x;k),$ in the region $x\le a$. Considering Exps. (\ref{500})
for the elements $q$ and $p$, we have

\begin{eqnarray} \label{266}
\Psi_{ref}(x;k)=-2\sqrt{R}\Api\sin\Big(k(x-a)+\frac{1}{2}
\left(\lambda-J+F-\frac{\pi}{2}\right)\Big) e^{i\phi_{(+)}}
\end{eqnarray}
where
\[\phi_{(\pm)}=\frac{1}{2}\left[\lambda \pm \left(J-F-\frac{\pi}{2}
+2ka\right) \right].\]

Now we have to show that only one of these solutions describes the
state of the subensemble of reflected particles. To select it, we have
to study both the solutions in the region $x\ge b$ where they can be
written in the form

\begin{eqnarray} \label{267}
\Psi_{ref}(x;k)=\Api(k)\left(\Apo^{ref}(k)e^{ikx}+
\Ami^{ref}(k)e^{-ikx}\right)
\end{eqnarray}
where
\[\Apo^{ref}=\sqrt{R} G^* e^{i\phi_{(+)}}; \ppp
\Ami^{ref}=\sqrt{R} G e^{i\phi_{(+)}},\ppp
G=q e^{-i\phi_{(-)}}-p^* e^{i\phi_{(-)}}.\]
Considering Exps. (\ref{500}) as well as the equality
$\exp(i\lambda)=\sqrt{R}\pm i\sqrt{T}$, one can show that
\[G=\mp i\exp\left[i\left(kb-\frac{1}{2} \left(J+F+\frac{\pi}{2}-\lambda
\right)\right) \right];\] here the signs ($\mp$) correspond to those in
the expression for $\lambda$.  Then, for $x\ge b$, we have

\begin{eqnarray} \label{268}
\Psi_{ref}(x;k)=\mp 2\sqrt{R}\Api\sin\Big[k(x-b)+\frac{1}{2}
\left(J+F+\frac{\pi}{2}-\lambda\right)\Big] e^{i\phi_{(+)}}.
\end{eqnarray}
For the following it is convenient to go over to the variable $x'$:
$x=x_{mid}+x'$ where $x_{mid}=(a+b)/2.$ Then we have, for $x'\le -d/2$,

\[\Psi_{ref}(x')=-2\sqrt{R}\Api\sin\Big[\frac{1}{2}\big(kd+\lambda
-J-\frac{\pi}{2}\big)
+\frac{F}{2}+kx'\Big] e^{i\phi_{(+)}},\]
for $x'\ge d/2$ ---
\[\Psi_{ref}(x')=\pm 2\sqrt{R}\Api\sin\Big[\frac{1}{2}\big(kd+ \lambda
-J-\frac{\pi}{2}\big)-\frac{F}{2}-kx'\Big]
e^{i\phi_{(+)}}.\]

From these expressions it follows that for any point $x'=x_0$ ($x_0\le
-d/2$) we have
\begin{eqnarray} \label{269}
\Psi_{ref}(x_0)=-2\sqrt{R}\Api\sin\Big[\frac{1}{2}\big(kd+ \lambda
-J-\frac{\pi}{2}+F\big)+k x_0\Big] e^{i\phi_{(+)}}
\end{eqnarray}

\begin{eqnarray} \label{270}
\Psi_{ref}(-x_0)=\pm 2\sqrt{R}\Api\sin\Big[\frac{1}{2}\big(kd+ \lambda
-J-\frac{\pi}{2}+F\big)+kx_0-F\Big] e^{i\phi_{(+)}}.
\end{eqnarray}

Let us consider the case of symmetric potential barriers:
$V(x')=V(-x')$. For such barriers the phase $F$ is equal to either 0 or
$\pi$. Then, as is seen from Exps. (\ref{269}) and (\ref{270}), one of
the above two stationary solutions $\Psi_{ref}(x';k)$ is odd in the
out-of-barrier region, but another function is even. Namely, when $F=0$
the upper sign in (\ref{270}) corresponds to the odd function, the
lower gives the even solution. On the contrary, when $F=\pi$ the second
root $\lambda$ leads to the odd function $\Psi_{ref}(x';k)$.

It is evident that in the case of symmetric barriers both the functions
keep their "out-of-barrier" symmetry in the barrier region as well.
Thus, the odd solution $\Psi_{ref}(x';k)$ is equal to zero at the point
$x'=0$. Of importance is the fact that this property takes place for
all values of $k$. In this case the probability flux, for any
nonstationary-state wave function formed only from the odd (or even)
stationary solutions $\Psi_{ref}(x';k)$, should be equal to zero at the
barrier's midpoint. This means that for particles impinging a symmetric
barrier from the left they are reflected by the barrier without
penetration into the region $x'\ge 0$. In its turn, this means that the
searched-for stationary-state RWF should be zero in the region $x'\ge
0$, but in the region $x'\le 0$ it must be equal to the odd function
$\Psi_{ref}(x';k)$. In this case the corresponding probability density
is everywhere continuous, including the point $x'=0$, and the
probability flux is everywhere equal to zero.

Of importance is the fact that the above property of reflection admits,
in principle, experimental checking. Indeed, since reflected particles
does not penetrate into the region $x\ge x_{mid}$ of the symmetric
barrier, the switching on an infinitesimal magnetic field in this
region must not influence the spin of these particles. For checking
this property, one can use the experimental scheme presented in
\cite{But}.

As regards the searched-for TWF, $\Psi_{tr}(x;k)$, it can be found now
from the expression $\Psi_{tr}(x;k)=\Psi_{full}(x;k)-\Psi_{ref}(x;k)$.
This function is everywhere continuous, and the corresponding
probability flux is everywhere constant (we have to stress once more
that this quantity has no discontinuity at the point $x=x_{mid}$,
though the first derivative of $\Psi_{tr}(x;k)$ is discontinuous at
this point). Thus, as in the case of the RWF, wave packets formed from
the stationary-state TWF should evolve in time with a constant norm.

As is seen from Exps. (\ref{269}) and (\ref{270}), for asymmetric
potential barriers, both the solutions $\Psi_{ref}(x';k)$ are neither
even nor odd functions. Nevertheless, it is evident that for any given
value of $k$ one of these solutions has opposite signs at the barrier's
boundaries. This means that, for any $k$, there is at least one point
in the barrier region, at which this function is equal to zero.
However, unlike the case of symmetric barriers, the location of such a
point depends on $k$. Therefore the behavior of the nonstationary-state
RWF in the barrier region is more complicated for asymmetric barriers.
Now the most right turning point for reflected particles lies, as in
the case of symmetric barriers, in the barrier region, but this point
does not coincide in the general case with the midpoint of this region.

To illustrate the temporal behavior of all the three wave functions,
i.e., $\Psi_{full}$, $\Psi_{tr}$ and $\Psi_{ref},$ we have considered
the case of rectangular barriers. In this case, the stationary-state
wave function $\Psi_{ref}(x;k)$, for $a\le x\le x_{mid}$, reads as

\begin{eqnarray} \label{271}
\Psi_{ref}=2\sqrt{R}\Api
e^{i\phi_{(+)}}\big[\cos(ka+\phi_{(-)})\sinh(\kappa d/2)\nonumber\\
-\frac{k}{\kappa}\sin(ka+\phi_{(-)})\cosh(\kappa
d/2)\big]\sinh(\kappa(x-x_{mid})) \end{eqnarray} where
$\kappa=\sqrt{2m(v_0-E)}/\hbar$ (the below-barrier case); and

\begin{eqnarray} \label{272}
\Psi_{ref}=-2\sqrt{R}\Api
e^{i\phi_{(+)}}\big[\cos(ka+\phi_{(-)})\sin(\kappa
d/2)\nonumber\\+\frac{k}{\kappa}\sin(ka+\phi_{(-)})\cos(\kappa
d/2)\big]\sin(\kappa(x-x_{mid}))
\end{eqnarray}
where $\kappa=\sqrt{2m(E-v_0)}/\hbar$ (the above-barrier case). In
both cases $\Psi_{ref}(x;k)\equiv 0$ for $x\ge x_{mid}.$

We have calculated the spatial dependence of the probability densities
$|\Psi_{full}(x,t)|^2$ (dashed line), $|\Psi_{tr}(x,t)|^2$ (open
circles) and $|\Psi_{ref}(x,t)|^2$ (solid line) for the rectangular
barrier ($V_0=0.3 eV$, $a=500 nm$, $b=505 nm$) and well ($V_0=-0.3 eV$,
$a=500 nm$, $b=505 nm$). Figures 1 ($t=0$), 2 ($t=0.4 ps$) and 3
($t=0.42 ps$) display results for the barrier, and figures 4 ($t=0$), 5
($t=0.4 ps$) and 6 ($t=0.43 ps$) display results for the well. In both
the cases, the function $\Psi_{full}(x,0)$ represents the Gaussian wave
packet with $l_0=7.5 nm$; the average kinetic energy is equal to $0.25
eV,$ both for the barrier and well. Besides, in both cases, the
particle's mass is $0.067 m_e$ where $m_e$ is the mass of an electron.

As is seen from figures 1 and 4, the average starting points for the
RWF and TWF differ from that for $\Psi_{full}$. The main peculiarity of
the transmitting wave packet is that it is slightly compressed in the
region of the barrier, and stretched in the region of the well. Figure
7 shows that, at the stage of the scattering event ($t=0.4 ps$; see
also figure 2), the probability to find a transmitting particle in the
barrier region is larger than in the neighborhood of the barrier. This
means that in the momentum space this packet becomes wider when the
ensemble of particles enters the barrier region. For the well (see
figure 8) there is an opposite tendency.  Note that for the barrier
$<T>_{in}\approx 0.149$. For the well $<T>_{in}\approx 0.863$.

\subsection {Connection of the wave functions for reflection and
transmission with the eigenvectors of the scattering matrix} \label{a23}

Of importance is the fact that there are other two settings of the
tunneling problem for the given potential $V(x)$ when the subensemble's
states described by the RWF and TWF arise explicitly. Indeed, let us
find such solutions to the Schr\"odinger equation, for a given
potential $V(x)$, for which
\begin{eqnarray} \label{600}
\left(\begin{array}{c} \bpo \\ \bmo \end{array} \right)=S
\left(\begin{array}{c} \bpi\\ \bmi
\end{array} \right)
\end{eqnarray}
where $S$ is a constant. This means that the amplitudes of incoming
waves should obey the characteristic equation

\begin{eqnarray} \label{601}
{\bf S} \left(\begin{array}{c} \bpi \\ \bmi
\end{array} \right)=S \left(\begin{array}{c} \bpi\\ \bmi
\end{array} \right); \ppp
{\bf S}=\left(\begin{array}{cc} q^{-1} & -p/q \\
p^*/q & q^{-1}
\end{array} \right)
\end{eqnarray}
where ${\bf S}$ is the scattering matrix.

It is easy to show that the solutions of this equation can be written
in the form

\[S=\frac{1+i\mu |p|}{q}; \ppp \left(\begin{array}{c} \bpi \\ \bmi
\end{array} \right) = c_{(\mu)} \left(\begin{array}{c} i\mu p/|p| \\ 1
\end{array} \right)\]
where $\mu=\pm 1;$ $c_{(+)}$ and $c_{(-)}$ are arbitrary constants.

Now let us find such values of $c_{(+)}$ and $c_{(-)}$ at which
$\bmo=p^*/q$. It easy to show that all four amplitudes read, in this
case, as
\begin{eqnarray} \label{602}
\bpi=\frac{i\mu |p|}{1+i\mu |p|}\equiv\sqrt{R}(\sqrt{R}+i\mu \sqrt{T});
\ppp\bmo=\frac{p^*}{q};\nonumber\\\bpo=\frac{i\mu |p|}{q} \equiv
\frac{i\mu |p|}{p} \cdot \frac{p}{q};\ppp\bmi= \frac{p^*}{1+i\mu
|p|}\equiv \frac{p^*}{i\mu |p|}\sqrt{R}(\sqrt{R}+i\mu \sqrt{T}).
\end{eqnarray}
One of two solutions with these amplitudes is evident to coincide, for
$x<a$, with the RWF found in subsection \ref{a22}. This means that in
the case of symmetric potential barriers this function, like the RWF,
is equal to zero at the midpoint of the barrier region, for any value
of $k$. In this two-sources scattering problem, both the incident wave
packets does not cross the above point. In fact, we deal here with the
ideal bilateral reflection of particles from the midpoint of the
barrier region, which is described by the sum of two the RWFs.

In a similar way, for the same eigenvalue of the scattering matrix, one
can find such values of $c_{(+)}$ and $c_{(-)}$ at which $\bpo=1/q$:

\begin{eqnarray} \label{603}
\bpi=\frac{1}{1+i\mu |p|}\equiv\sqrt{T}(\sqrt{T}-i\mu \sqrt{R});
\ppp\bmo=-\frac{i\mu |p|}{p}\cdot\frac{1}{q};\nonumber\\
\bpo=\frac{1}{q}; \ppp\bmi=-\frac{i\mu |p|}{p}
\frac{1}{1+i\mu |p|}\equiv -\frac{i\mu |p|}{p}\cdot
\sqrt{T}(\sqrt{T}-i\mu \sqrt{R}).
\end{eqnarray}

As is seen, the stationary-state TWF appears explicitly in the solution
with amplitudes (\ref{603}). In the case of symmetric potential
barriers this solution is evident to represent a sum of two continuous
wave functions whose probability fluxes are continuous too. For one of
them the amplitudes of incoming and outgoing waves are, respectively,
$\bpi$ ($=\sqrt{T}(\sqrt{T}-i\mu \sqrt{R})$) and $\bpo$ ($=1/q$). For
another function these amplitudes are, respectively, $\bmi$ ($= -(i\mu
|p|/p)\cdot \sqrt{T}(\sqrt{T}-i\mu \sqrt{R})$) and $\bmo$ ($= -(i\mu
|p|/p)\cdot (1/q)$). The first (second) function is just the TWF to
describe the ideal transmission of particles impinging the barrier from
the left (right). The corresponding nonstationary-state wave functions
are evident to evolve in time with a constant norm.

So, each of the above "two-sources" wave functions generated by
eigenvectors of the scattering matrix represent a sum of two causally
evolved "one-source" wave functions. One of them describes the state of
a particle impinging the barrier from the left. Another function
relates to particles moving to the right of the barrier. In the case of
(\ref{602}) both the one-source wave packets are ideally reflected by
the barrier. And, in the case of (\ref{603}) both one-source wave
packets are ideally transmitted by it. These two auxiliary tunneling
problems give us the basis to verify the formalism presented in this
paper.

Note also that the stationary-state RWF and TWF, for the problem at
hand, should correspond to the same value of $\mu$, i.e., to the same
eigenvalue of the scattering matrix.  As regards another eigenvalue, in
the case of reflection it generates the even function which does not
fit as a RWF (see subsection \ref{a22}). That is, only one of the
eigenvalues of the scattering matrix is associated with the RWF and
TWF of the scattering problem considered.

\section{Exact and asymptotic tunneling times for transmission and
reflection} \label{a3}
\subsection{Exact tunneling times} \label{a31}

\hspace*{\parindent} So, we have found two causally evolved wave
packets to describe the subensembles of transmitted and reflected
particles in the considered tunneling problem, at all stages of the
scattering process. As is shown, the motion of these packets can be, in
principle, observed experimentally. It is evident that the given
formalism may serve as the basis to solve the tunneling time problem,
since now one can follow the CMs of wave packets, which describe
separately reflection and transmission, at all instants of time.

Let $t^{tr}_1$ and $t^{tr}_2$ be such instants of time that

\begin{equation} \label{80}
\frac{<\Psi_{tr}(x,t^{tr}_1)|\hat{x}|\Psi_{tr}(x,t^{tr}_1)>}
{<\Psi_{tr}(x,t^{tr}_1)|\Psi_{tr}(x,t^{tr}_1)>} =a-L_1;
\end{equation}
\begin{equation} \label{81}
\frac{<\Psi_{tr}(x,t^{tr}_2)|\hat{x}|\Psi_{tr}(x,t^{tr}_2)>}
{<\Psi_{tr}(x,t^{tr}_1)|\Psi_{tr}(x,t^{tr}_1)>} =b+L_2,
\end{equation}

\noindent where $\Psi_{tr}(x,t)$ is the subensemble's wave function
found above for transmission. Then, one can define the transmission
time $\Delta t_{tr}(L_1,L_2)$ as the difference $t^{tr}_2(L_2)-
t^{tr}_1(L_1)$ where $t^{tr}_1(L_1)$ is the smallest root of Eq.
(\ref{80}), and $t^{tr}_2(L_2)$ is the largest root of Eq.  (\ref{81}).

Similarly, for reflection, let $t^{ref}_1(L_1)$ and $t^{ref}_2(L_1)$ be
such instants of time $t$ that

\begin{equation} \label{110}
\frac{<\Psi_{ref}(x,t)|\hat{x}|\Psi_{ref}(x,t)>}
{<\Psi_{ref}(x,t)|\Psi_{ref}(x,t)>}=a-L_1,
\end{equation}

\noindent Then the reflection time $\Delta t_{ref}(L_1)$ can be defined
as $\Delta t_{ref}(L_1)=t^{ref}_2-t^{ref}_1$ where $t^{ref}_1(L_1)$ is
the smallest root, and $t^{ref}_2(L_2)$ is the largest root of Eq.
(\ref{110}) (of course, if they exist).

It is important to emphasize that, due to conserving the number of
particles in both the subensembles, both these quantities are
non-negative for any distances $L_1$ and $L_2$. Both the definitions
are valid, in particular, when $L_1=0$ and $L_2=0$. In this case the
quantities $\Delta t_{tr}(0,0)$ and $\Delta t_{ref}(0)$ yield,
respectively, exact transmission and reflection times for the barrier
region. Of course, one has to take into account that in the case of
reflection the CM of the wave packet may turn back without entering the
barrier region.

\newcommand {\uta} {\tau_{tr}}
\newcommand {\utb} {\tau_{ref}}

\subsection{Asymptotic tunneling times}

It is evident that in the general case the above average quantities can
be calculated only numerically. At the sane time, for sufficiently
large values of $L_1$ and $L_2$, one can obtain the tunneling times
$\Delta t_{tr}(L_1,L_2)$ and $\Delta t_{tr}(L_1,L_2)$ in more explicit
form. Indeed, in this case, instead of the exact subensemble's wave
functions, we can use the corresponding in asymptotes derived in
$k$-representation. Indeed, now the "full" in asymptote, like the
corresponding out asymptote, represents the sum of two wave packets:
\[f_{in}(k,t)=f_{in}^{tr}(k,t)+f_{in}^{ref}(k,t);\]
\begin{eqnarray} \label{75}
f^{tr}_{in}(k,t)=\sqrt{T(k)}\Api(k)\exp[i(\Lambda(k)
-\alpha\frac{\pi}{2} - E(k)t/\hbar)];
\end{eqnarray}
\begin{eqnarray} \label{76}
f^{ref}_{in}(k,t)=\sqrt{R(k)}\Api(k)\exp[i(\Lambda(k)- E(k)t/\hbar)];
\end{eqnarray}
$\alpha=1$ if $\Lambda\ge 0$; otherwise $\alpha=-1.$
Here the function $\Lambda(k)$ coincides, for a given $k$, with one of
the functions, $\lambda(k)$ or $-\lambda(k)$, for which
$\Psi_{ref}(x;k)$ is an odd function (see above). One can easily show
that for both the roots
\[|\Lambda^\prime(k)|=\frac{|T^\prime|}{\sqrt{2 R T}}.\]

A simple analysis in the $k$-representation shows that well before the
scattering event the average kinetic energy of particles in both
subensembles (with the average wave numbers $<k>^{tr}_{in}$ and
$<k>^{ref}_{in}$) is equal to that for large times:
\[<k>^{tr}_{out}=<k>^{tr}_{in}, \ppp <k>^{ref}_{out}=-<k>^{ref}_{in}.\]
Besides, at early times

\begin{eqnarray} \label{73}
<\hat{x}>^{tr}_{in}=\frac{\hbar t}{m}<k>^{tr}_{in}
-<\Lambda^\prime(k)>^{tr}_{in};
\end{eqnarray}

\begin{eqnarray} \label{74}
<\hat{x}>^{ref}_{in}=\frac{\hbar t}{m}<k>^{ref}_{in}
-<\Lambda^\prime(k)>^{ref}_{in}
\end{eqnarray}

As it follows from Exps. (\ref{73}) and (\ref{74}), the average
starting points $x_{start}^{tr}$ and $x_{start}^{ref}$, for the
subensembles of transmitted and reflected particles, respectively,
differ from that for all particles:
\begin{eqnarray} \label{730}
x_{start}^{tr}=-<\Lambda^\prime(k)>^{tr}_{in},\ppp x_{start}^{ref}=
-<\Lambda^\prime(k)>^{ref}_{in}.
\end{eqnarray}
The implicit assumption made in the SWPA that incident, as well as
transmitted and reflected particles start, on the average, from the
same point does not agree with this result. By our approach, this is
the main reason why the asymptotic transmission and reflection times
obtained in the SWPA should be considered as ill-defined quantities,
for any wave packets.

Let us take into account Exps. (\ref{73}), (\ref{74}) and again
analyze the motion of a particle in the above spatial interval covering
the barrier region.  In particular, let us calculate the transmission
time, $\uta$, spent (on the average) by a particle in the interval
$[Z_1, Z_2]$. It is evident that the above equations for the arrival
times $t^{tr}_1$ and $t^{tr}_2$, which correspond the extreme points
$Z_1$ and $Z_2$, respectively, read now as

\[<\hat{x}>^{tr}_{in}(t^{tr}_1)=a-L_1;
\ppp<\hat{x}>^{tr}_{out}(t^{tr}_2)=b+L_2.
\]

\noindent Considering (\ref{73}) and (\ref{64}), we obtain from here
that now the transmission time is
\begin{eqnarray} \label{23}
\uta(L_1,L_2)\equiv t^{tr}_2-t^{tr}_1=\frac{m}{\hbar
<k>^{tr}_{in}}\left(<J^\prime>^{tr}_{out}
-<\Lambda^\prime>^{tr}_{in} +L_1+L_2 \right).
\end{eqnarray}

Similarly, for the reflection time, $\utb(L_1)$
($\utb=t^{ref}_2-t^{ref}_1$), we have

\[<\hat{x}>^{ref}_{in}(t^{ref}_1)=a-L_1,
\ppp<\hat{x}>^{ref}_{out}(t^{ref}_2)=a-L_1.
\]
\noindent Considering (\ref{74}) and (\ref{65}), one can easily show
that

\begin{eqnarray} \label{25}
\utb(L_1)\equiv t^{ref}_2-t^{ref}_1=\frac{m}{\hbar
<k>^{ref}_{in}}\left(<J^\prime -
F^\prime>^{ref}_{out}-<\Lambda^\prime>^{ref}_{in} +2L_1\right).
\end{eqnarray}

The inputs $\uta^{as}$ ($\uta^{as}=\uta(0,0)$) and $\utb^{as}$
($\uta^{as}=\uta(0,0)$) will be named below as the asymptotic
transmission and reflection times for the barrier region, respectively:

\begin{eqnarray} \label{230}
\uta^{as}=\frac{m}{\hbar <k>^{tr}_{in}}\Big(<J^\prime>^{tr}_{out}
-<\Lambda^\prime>^{tr}_{in}\Big),
\end{eqnarray}
\begin{equation} \label{250}
\utb^{as}=\frac{m}{\hbar <k>^{ref}_{in}}\left(<J^\prime -
F^\prime>^{ref}_{out}-<\Lambda^\prime>^{ref}_{in}\right)
\end{equation}
Here the word "asymptotic" points to the fact that these quantities were
obtained with making use of the in and out asymptotes for the
subensembles investigated. Unlike the exact tunneling times the
asymptotic times may be negative by value.

The corresponding lengths $d_{eff}^{tr}$ and $d_{eff}^{ref},$

\begin{eqnarray} \label{251}
d_{eff}^{tr}=<J^\prime>^{tr}_{out}
-<\Lambda^\prime>^{tr}_{in},\ppp
d_{eff}^{ref}=<J^\prime-F^\prime>^{ref}_{out}
-<\Lambda^\prime>^{ref}_{in},
\end{eqnarray}
can be treated as the effective widths of the barrier for transmission
and reflection, respectively.

\subsection{Average starting points and asymptotic tunneling times for
rectangular potential barriers and $\delta$-potentials} \label{a33}

Let us consider the case of a rectangular barrier (or well) of height
$V_0$ and obtain explicit expressions for $d_{eff}(k)$
(now, both for transmission and reflection, $d_{eff}(k)=J^\prime(k)
-\Lambda^\prime(k)$ since $F^\prime(k)\equiv 0$) which can be treated
as the effective width of the barrier for a particle with a given $k$.
Besides, we will obtain the corresponding expressions for the
coordinate, $x_{start}(k)$, of the average staring point for this
particle:  $x_{start}(k)=-\Lambda^\prime(k)$. It is evident that in
terms of $d_{eff}$ the above asymptotic times for a particle with the
well-defined average momentum $k_0$ read as
\[\uta^{as}=\utb^{as}=\frac{m d_{eff}(k_0)}{\hbar k_0}.\]

Using the expressions for the real tunneling parameters $J$ and $T$
(see \cite{Ch1,Ch4}), one can show that, for the below-barrier case
($E\le V_0$),

\[d_{eff}(k)=\frac{4}{\kappa}
\frac{\left[k^2+\kappa_0^2\sinh^2\left(\kappa d/2\right)\right]
\left[\kappa_0^2\sinh(\kappa d)-k^2 \kappa d\right]}
{4k^2\kappa^2+ \kappa_0^4\sinh^2(\kappa d)}\]

\[x_{start}(k)= -2\frac{\kappa_0^2}{\kappa}
\frac{(\kappa^2-k^2)\sinh(\kappa d)+k^2 \kappa d
\cosh(\kappa d)} {4k^2\kappa^2+ \kappa_0^4\sinh^2(\kappa d)}\]
where $\kappa=\sqrt{2m(V_0-E)/\hbar^2};$
for the above-barrier case ($E\ge V_0)$ ---
\[d_{eff}(k)=\frac{4}{\kappa} \frac{\left[k^2-\beta
\kappa_0^2\sin^2\left(\kappa d/2\right)\right]\left[k^2 \kappa d-\beta
\kappa_0^2\sin(\kappa d)\right]} {4k^2\kappa^2+\kappa_0^4\sin^2(\kappa
d)}\]

\[x_{start}(k)= -2\beta \frac{\kappa_0^2}{\kappa} \cdot
\frac{(\kappa^2+k^2)\sin(\kappa d)-k^2 \kappa d
\cos(\kappa d)} {4k^2\kappa^2+ \kappa_0^4\sin^2(\kappa d)}\]
where
$\kappa=\sqrt{2m(E-V_0)/\hbar^2};$ $\beta=1$ if $V_0>0$, otherwise,
$\beta=-1$. In both the cases $\kappa_0=\sqrt{2m|V_0|/\hbar^2}$.

It is important to stress that, in the limit $k\to \infty,$ $d_{eff}\to
d$ and $x_{start}(k) \to 0$. This property guarantees that for
infinitely narrow in $x$-space wave packets the average starting points
for both subensembles will coincide with that for all particles.  It is
important also that for wells the values of $d_{eff}$ and, as a
consequence, the corresponding tunneling times are negative, in the
limit $k\to 0$, when $\sin(\kappa_0 d)<0$.

Note that for sufficiently narrow barriers and wells, namely when
$\kappa d\ll 1$, we have $d_{eff}\approx d$. That is, particles tunnel,
on the average, classically through such barriers. For the starting
point we have
\[x_{start}(k)\approx -\frac{\kappa_0^2}{2 k^2} d, \ppp
x_{start}(k)\approx -\beta \frac{\kappa_0^2}{2 k^2} d\]
for $E\le V_0$ and $E\ge V_0$, respectively.

For wide barriers and wells, when  $\kappa d\gg 1$, we have
$d_{eff}\approx 2/\kappa$ and $x_{start}(k)\approx 0$, for $E\le
V_0$; and
\[d_{eff}\approx 4k^2 d \cdot \frac{k^2-\beta\kappa_0^2 \sin^2(\kappa
d/2)}{4k^2\kappa^2+\kappa_0^4 \sin^2(\kappa d)}\ppp
x_{start}(k)\approx  \frac{2\beta\kappa_0^2 k^2 d \cos(\kappa d)}{4
k^2 \kappa^2+ \kappa_0^4 \sin^2(\kappa d)},\] for $E\ge V_0$.

It is interesting to note that for the $\delta$-potential $V(x)=
W \delta(x-a)$ $d_{eff}\equiv 0$. This means that, contrary to the
phase tunneling time, the tunneling times defined here equal to zero
for this potential. As regards the starting point $x_{start}(k)$ in the
case of the $\delta$-potential, we have \[x_{start}(k)=-\frac{2m\hbar^2
W}{\hbar^4 k^2+m^2W^2}.\] Thus, we see that, for example, in the case
of $\delta$-wells ($W<0$) particles in each subensemble start, on the
average, with an advance in comparison with those of the whole quantum
ensemble.

\section{Conclusion}

A separate description of transmission and reflection is commonly
accepted to contradict the principles of quantum mechanics. However, in
this paper we argue that this is not the case, at least in the
one-dimensional one-particle scattering problem. We show that the wave
function to describe, in this problem, the state of the whole ensemble
of identically prepared particles can be uniquely presented as the sum
of two functions (named here as the TWF and RWF) to obey the
Schr\"odinger equation. In the case of nonstationary case, these
functions are mutually orthogonal. At late times the TWF coincides
with the transmitted wave packet, and the RWF approaches the reflected
one. We postulate that namely the TWF and RWF are the wave functions to
describe, respectively, transmission and reflection in the considered
scattering process.

There is also a widely accepted viewpoint (see, for example, page 106 in
\cite{Bal} and page 17 in \cite{Lan}) that all solutions to the
stationary one-dimensional Schr\"odinger equation, for a finite
potential, must be everywhere continuous together with their spatial
derivatives; otherwise, the points where this requirement is violated
contain allegedly sinks or sources of particles.  However, in this
paper we show that the above requirement for "physical" solutions to
the Schr\"odinger equation is, in reality, excessive. The main
peculiarity of the presented stationary-state wave functions for
transmission and reflection is that there is a point in the barrier
region where these everywhere continuous functions have discontinuous
first spatial derivatives (in the case of symmetric potential barriers
this takes place at the midpoint of the barriers). Nevertheless, for
each subensemble, this point contains neither sink nor source of
particles:  both for transmission and reflection, the probability
current density for each stationary-state wave function is constant on
the spatial axis, and the norm of wave packets formed from these
functions is constant in time.

We show that, in the case of a symmetric potential barrier, reflected
particles impinging the barrier from the left do not penetrate into the
spatial domain lying to the right of the midpoint of the barrier
region. This means, in particular, that the switching on an
infinitesimal magnetic field in this domain must not influence the spin
of these particles.

Besides, for the given potential we formulate two scattering problem in
which the RWF and TWF arise separately and, as a consequence, there is
another possibility to check experimentally our approach. In both the
scattering problems the amplitudes of incoming waves form the
eigenvectors of the scattering matrix for the given potential.

On the basis of the above formalism we define average (exact and
asymptotic) transmission and reflection times. The exact tunneling
times are always non-negative. In the case of rectangular barriers and
$\delta$-potentials, for both the subensembles, we derive explicit
expressions for the asymptotic tunneling times and for the average
starting points. These times differ essentially from those arising in
the SWPA.

 \newpage
\section*{Figure captions}

Fig. 1 The $x$-dependence of $|\Psi_{full}(x,t)|^2$ (dashed line) which
represents the Gaussian wave packet with $l_0\approx 7.5 nm$ and
the average kinetic particle's energy $0.25 eV$, as well as
$|\Psi_{tr}(x,t)|^2$ (open circles) and $|\Psi_{ref}(x,t)|^2$ (solid
line) for the rectangular barrier ($V_0=0.3 eV$, $a=500 nm$, $b=505
nm$); $t=0$.

\vspace{1cm}
\noindent Fig. 2 The same as in Fig. 1, but $t=0.4 ps$.

\vspace{1cm}
\noindent Fig. 3 The same as in Fig. 1, but $t=0.42 ps$.

\vspace{1cm}
\noindent Fig. 4 The $x$-dependence of $|\Psi_{full}(x,t)|^2$ (dashed
line) which represents the Gaussian wave packet with $l_0=7.5 nm$ and
the average kinetic particle's energy $0.25 eV$, as well as
$|\Psi_{tr}(x,t)|^2$ (open circles) and $|\Psi_{ref}(x,t)|^2$ (solid
line) for the rectangular well ($V_0=-0.3 eV$, $a=500 nm$, $b=505 nm$);
$t=0$.

\vspace{1cm}
\noindent Fig. 5 The same as in Fig. 4, but $t=0.4 ps$.

\vspace{1cm}
\noindent Fig. 6 The same as in Fig. 4, but $t=0.43 ps$.

\vspace{1cm}
\noindent Fig. 7 The same functions for the barrier region;
parameters are the same as for Fig. 2.

\vspace{1cm}
\noindent Fig. 8 The same functions for the barrier region;
parameters are the same as for Fig. 5.

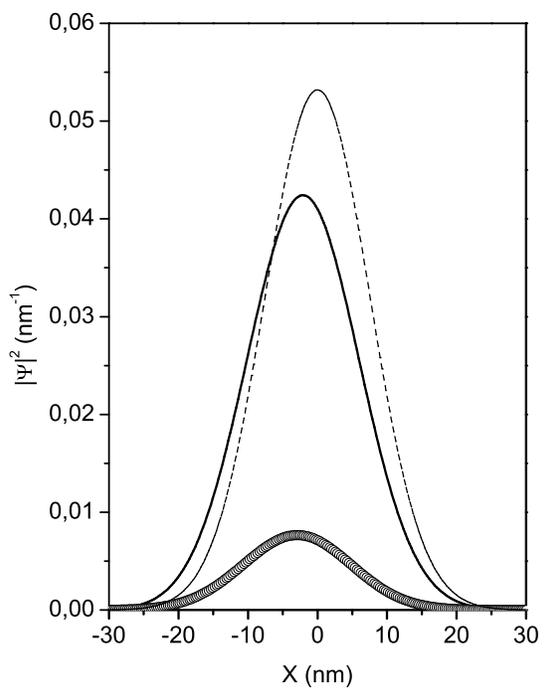
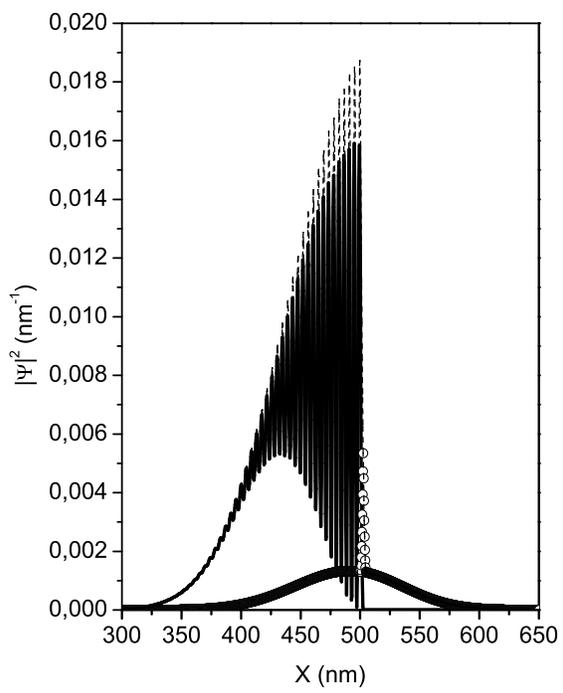
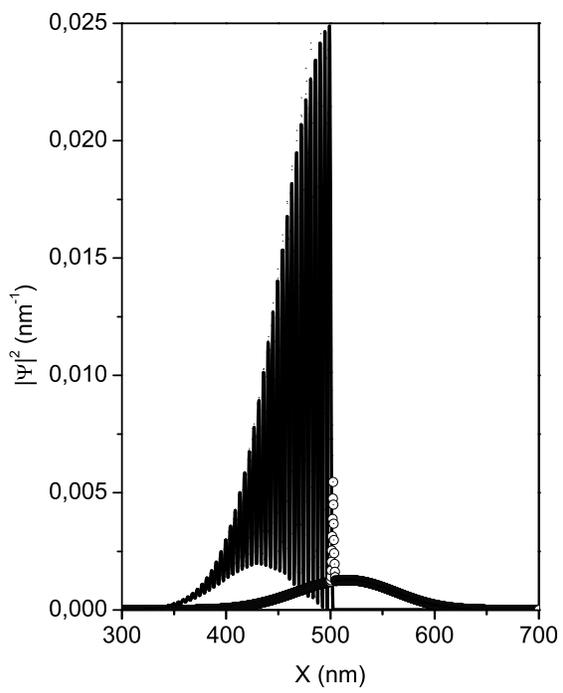
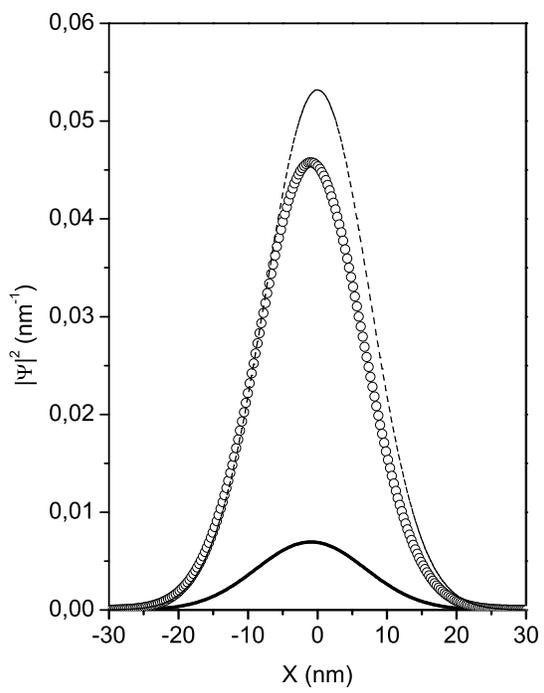
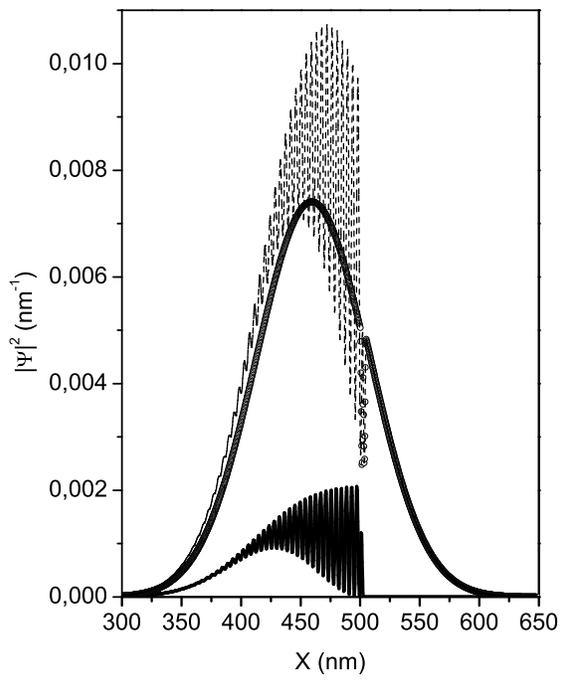
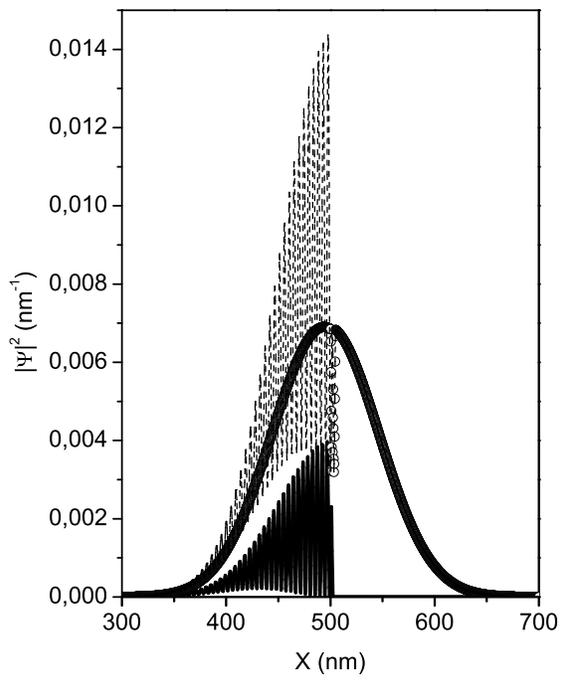
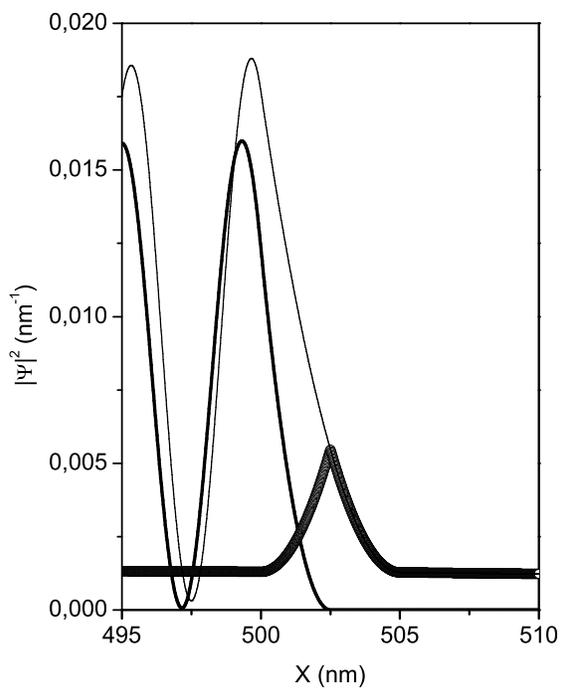
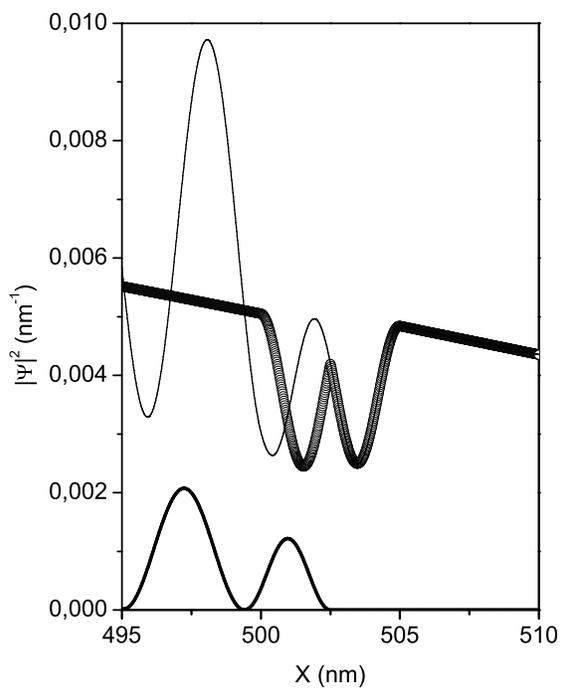
\end{document}